# Perceived Usability of Collaborative Modeling Tools

**Ranci Ren[1], John W. Castro[2*], Santiago R. Acuña[3], Oscar Dieste[4], Silvia T. Acuña[1]**

[1]Departamento de Ingeniería Informática, Universidad Autónoma de Madrid, Calle Francisco Tomás y Valiente 11, C.P. 28049, Campus Cantoblanco Madrid, Spain.

[2]Departamento de Ingeniería Informática y Ciencias de la Computación, Universidad de Atacama, Av. Copayapu 485, C.P. 1530000, Copiapó, Chile.

[3]Universidad Autónoma de San Luis Potosí, Álvaro Obregón 64, Col. Centro, C.P. 78000 San Luis Potosí, Mexico.

[4]Lenguajes y Sistemas Informáticos e Ingeniería de Software, Universidad Politécnica de Madrid, Calle de los Ciruelos, C.P. 28660, Boadilla del Monte, Spain.

[1]ranci.ren@uam.es, [2]john.castro@uda.cl, [3]santiago.acuna@uaslp.mx, [4]odieste@fi.upm.es, [1]silvia.acunna@uam.es

**ABSTRACT**

**Context:** Online collaborative creation of models is becoming commonplace. Collaborative modeling using chatbots and natural language may lower the barriers to modeling for users from different domains. **Objective:** We compare the perceived usability of two similarly online collaborative modeling tools, the SOCIO chatbot and the Creately web-based tool. **Method:** We conducted a crossover experiment with 66 participants. The evaluation instrument was based on the System Usability Scale (SUS). We performed a quantitative and qualitative exploration, employing inferential statistics and thematic analysis. **Results:** The results indicate that chatbots enabling natural language communication enhance communication and collaboration efficiency and improve the user experience. **Conclusion:** Chatbots need to improve guidance and help for novices, but they appear beneficial for enhancing user experience.

## 1. Introduction

Collaborative modeling is relevant in software engineering because it supports early discussion and knowledge building. However, it requires advanced context awareness, communication, discussion, decision-making, and coordination tools (e.g., collaborative protocols) (Pérez-Soler et al., 2018).

The use of social networks for software engineering has been acknowledged as having a potentially significant impact on practice and tools (Begel et al., 2010) (Storey et al., 2010). One of the reasons is that they are an agile and lightweight means for coordination and information sharing. However, some challenges include increasing community and end-user involvement, enhancing project coordination, and improving development activities (Storey et al., 2010).

The SOCIO chatbot is a collaborative modeling tool in social networks (using the nickname @ModellingBot on Telegram). Users send natural language commands to SOCIO. It responds to the user's commands by building new elements, each highlighted in a different color than the existing elements. The approach is inherently collaborative and integrates with the social networks with which users are familiar, increasing class diagram construction's flexibility (Pérez-Soler et al., 2017, 2018). Chatbots are used in other domains, like healthcare and e-commerce, where they compete with GUI-based tools. Chatbot adoption requires seamless integration into the user work environment where GUIs are widespread nowadays.

The results of a previous systematic mapping study show that no other chatbots support modeling through social media (Ren et al., 2022). Notably, Ribeiro et al. (Ribeiro et al., 2022) developed a mediator bot that aggregated information from multiple bots into a single pull request comment. Then they evaluated newcomers' perceptions of the bot and user preferences by comparing it to multiple bots. However, their bot was designed to help find and process existing information from multi-bots rather than to assist in creating models. Because of this, we see the opportunity for the SOCIO chatbot to close this gap. We are interested in evaluating perceived chatbot usability. The usability evaluation identifies SOCIO's ability to facilitate collaboration, thus providing insights into improving its performance.

---

\* Corresponding Author.

We believe that SOCIO's strengths and weaknesses can be better understood if compared to a state-of-the-art tool in the same domain. As previously mentioned, there are no other chatbots that support collaborative modeling. However, Creately (https://creately.com/app) is a real-time collaborative modeling tool offering over 50 types of diagrams. It provides drag-and-drop functionality and the diagramming components needed to build class diagrams. Creately uses a traditional web-based graphical user interface (GUI). Creately performs real-time dragging and dropping actions in response to the user. Users can change the class diagram appearance collaboratively. It should be noted that Creately has real-time response function and collaborative creation function in class diagram construction, which are similar to the chatbot SOCIO.

Given that Creately is one of the few UML modeling tools which supports collaborative modeling (Lanubile et al., 2010) (Ozkaya, M., 2019), we adopted Creately as the control tool to perform a controlled experiment. We searched for previous usability evaluations of Creately before conducting our research; however, we failed to find any relevant studies.

The experiment has been designed as a within-subject crossover experiment. Each experimental team consisted of three members, and each team was required to implement two different class diagram construction tasks. This experiment builds upon previous research (Ren et al., 2020), adding a qualitative procedure (i.e., thematic analysis) to round out the usability assessment. We evaluate the SOCIO chatbot performance on collaborative modeling tasks.

The experimental results were quantitatively and qualitatively analyzed by adopting a modified SUS questionnaire and thematic analysis. On the one hand, the SUS score is higher for SOCIO than for Creately, indicating SOCIO quantitatively outperformed Creately in perceived usability. On the other hand, the thematic analysis indicates that although both tools receive negative reviews and require some improvements regarding providing guidance and help for novices, SOCIO appears to outperform Creately qualitatively, as it receives more positive and less negative feedback.

Our research shows that chatbots need to provide better guidance and help for novices. Still, they contribute to satisfaction and communication efficiency, improve the user experience, and reduce modeling barriers for users from different fields.

This paper is organized as follows. Section 2 describes the experimental design. Section 3 reports the statistical analysis. Section 4 describes the thematic analysis. Section 5 reports the participants' feedback. Finally, Section 6 outlines the conclusions.

## 2. Experiment Design
### 2.1. Objectives, Hypotheses, and Variables

The null and alternative hypotheses are:

H.0: There is no difference in perceived usability using SOCIO or Creately when building a class diagram.

H.1: There is a (two-tailed) difference in perceived usability using SOCIO or Creately when building a class diagram.

The main independent variable is the modeling tool, with either the SOCIO chatbot or the Creately online application (i.e., two tools with similar functions) as treatments. The experiment response variable is usability. We measure perceived usability as satisfaction using the System Usability Scale (SUS) questionnaire.

### 2.2. Subjects

We used convenience sampling to recruit 66 student participants from two universities in two countries: 54 from the Universidad de las Fuerzas Armadas (Ecuador) and 12 from the Universidad Autónoma de Madrid (Spain). All participants are undergraduate students who are pursuing their Computing degrees. None are researchers or practitioners. All participants volunteered to participate. The participants' performance in the experiment did not affect their academic attainment.

During recruitment, we asked participants if they had ever used either of the two tools, and we only included participants who claimed that they had not. We also checked if the emails they use most frequently were registered with Creately.

Based on the demographic survey, we collected basic information from all participants. They featured the following characteristics:

- The final sample consisted of 66 participants. Out of the sample, 54 are men, and 12 are women.

- Participants have a mean age of 22 years with a standard deviation of 1.68. The highest concentration of participants is within the range of 22 to 23 years old.
- 92.4% of participants use social media frequently. The social media most used by participants are WhatsApp, Facebook, Instagram, and Telegram.
- Participants were all acquainted with class diagrams and had sufficient English knowledge.
- 47% of participants have used or frequently used the Telegram application, while 13.6% have never used this app.
- 95.5% of participants believe they are relatively familiar with class diagrams.
- 27.1% of participants have never used a chatbot, whereas 41.4% have related experience (20.0% have used chatbots occasionally, and 21.4% are regular users).

### 2.3. Experimental Design

We used a two-sequence, two-period within-subjects crossover design (see Table 1) to increase statistical power (Ren et al., 2020). Participants were randomly assigned first to three-member teams (remember that the task is collaborative modeling) and then to experimental groups (Group 1 or Group 2).

Before the experiment, participants completed a demographic survey and a consent form. Then a 10-minute introductory session provided instructions on using the tool participants were to use in the following period. Next, participants were given 30 minutes to complete a task. Each team applied both treatments, SOCIO (SC) and Creately (CR), in a different order (SC-CR/CR-SC). After each treatment, participants were asked to complete a tailored System Usability Scale (SUS) questionnaire to assess the user experience quantitatively (Brooke, 1996; Bangor et al., 2009). SUS has been used to evaluate the usability of GUI-based systems and chatbots (Zwakman et al., 2020). Participants were asked which of the two tools they preferred in the second SUS questionnaire. Finally, we designed and implemented an open-ended questionnaire to assess prominent user experience themes: positive and negative aspects of the tools, suggestions on how to improve the tool, and user preferences.

Table 1. Experimental Design

| Group | Period 1 (Task 1) | Period 1 (Task 1) |
|---|---|---|
| Group 1 (SC-CR) | Group 1 (SC-CR) | Group 1 (SC-CR) |
| SOCIO | SOCIO | SOCIO |

### 2.4. Tasks

We designed two tasks: (1) develop a class diagram representing a store, including product and customer management, and (2) design a class diagram for a school containing courses and students. We counterbalanced the difficulty levels and adjusted the class diagram complexity according to the experiment length.

### 3. Statistical Analysis

The SUS team score was defined as the median of the scores given by all three team members. The maximum score for both tools is 100. Table 1 shows that SOCIO outperforms Creately on satisfaction (p=0.03) with a d=0.52 (medium) effect size. SUS scores (Table 1) appear to be higher (7.56 points) for participants using SOCIO than Creately.

Table 1. Linear mixed model for satisfaction

|  | Estimate | Std. Error | *p-value* |
|---|---|---|---|
| *(Intercept)* | 60.63 | 4.51 | 0 |
| *Seq.* | -0.17 | 5.56 | 0.98 |
| *Treatment* | 7.56 | 3.13 | **0.03** |
| *Order* | -1.31 | 3.13 | 0.68 |

The SOCIO chatbot usability scores ranged from 35 to 92.5, with a median of 67.4. The Creately web tool scores ranged from 20 to 82.5, with a median of 59.9. According to the official validation, SOCIO and

Creately are regarded as D (average) on a scale from F (failure) to A+ (outstanding) (Bangor et al., 2009). This suggests that both SOCIO and Creately are usable collaborative modeling tools with minor defects.

## 4. Thematic Analysis

Thematic analysis is a method whereby qualitative data can be identified, analyzed, and reported (Braun and Clarke, 2006). Note, however, that thematic analysis is a flexible method that places no limits on how the conclusions can be drawn. Since we set out to identify repeated themes to pinpoint interesting aspects, any opinion repeated more than three times was regarded as a common participant view. We coded features mentioned more than three times within the data set. We identified five features based on measures of satisfaction (Hornbæk, 2006) for both SOCIO and Creately: interface, communication, content, collaboration, and task (see Figure 1 and 2).

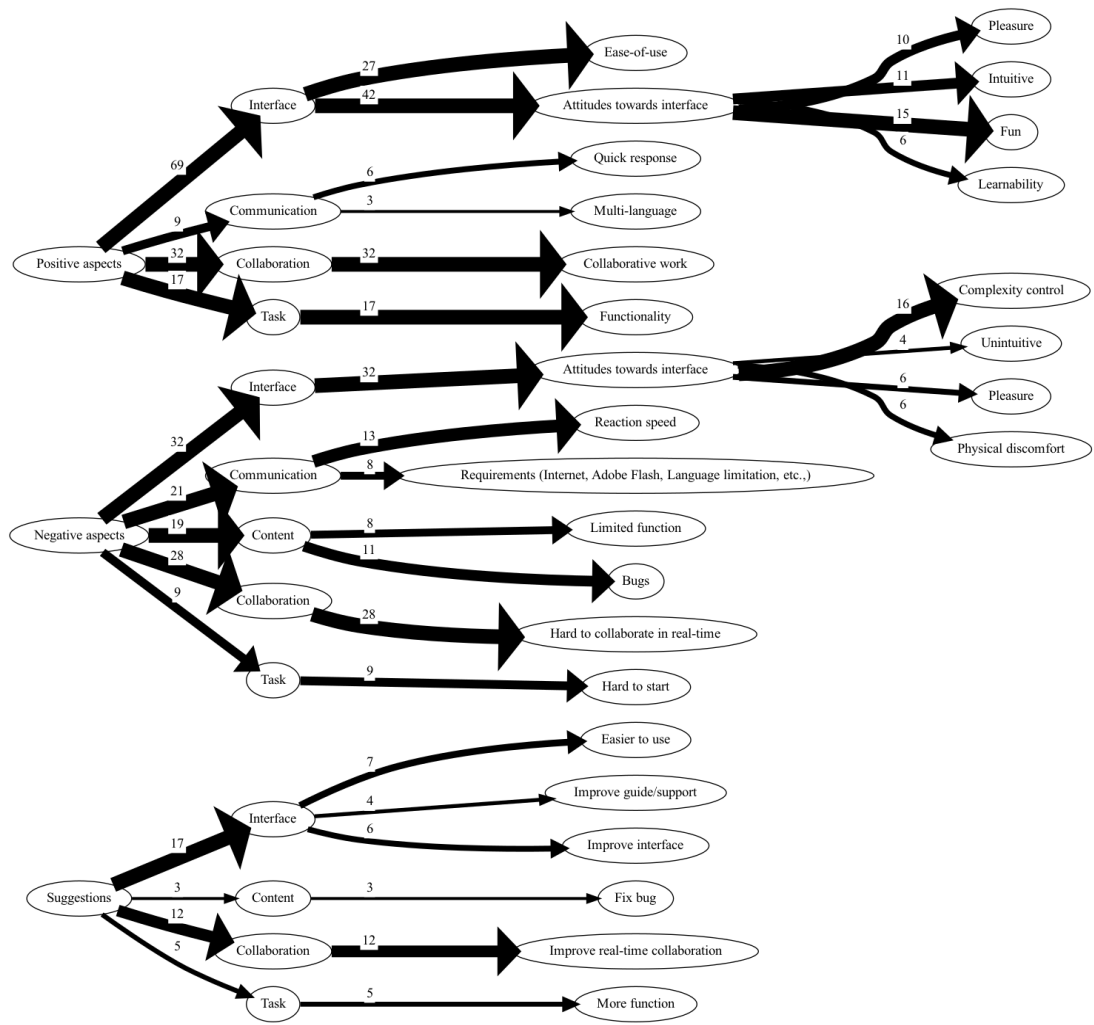

Figure 1. Thematic analysis graph for Creately

### 4.1. Interface

*Need for online support and help page for the modeling chatbot:* The online support and help page received a divisive assessment. Some participants (9) gave positive feedback about the help page because they found it integrative. Some (17) noted that the online support and help page was unsatisfactory because examples of complex diagrams and explanations of the related commands were poor.

Users that send wrong commands receive a message stating "I don't understand" without elaborating on where/what the error is. The SOCIO help page contains a four-minute video on how to create a simple class diagram and six graphs on how to create different elements of a class diagram. However, the SOCIO chatbot fails to provide intuitive and interactive instructional support for novice interactive interface users. Therefore, instead of receiving immediate support, users encountering problems during class diagram creation must read the text line by line or watch a video.

*Influence of attitude towards the interface:* The SOCIO chatbot offers users a more comprehensive range of humanized feelings. Some participants (10) considered SOCIO to be more friendly and reliable, as it enables more attractive, entertaining, and valid interaction (e.g., "it's so much fun collaborating like this").

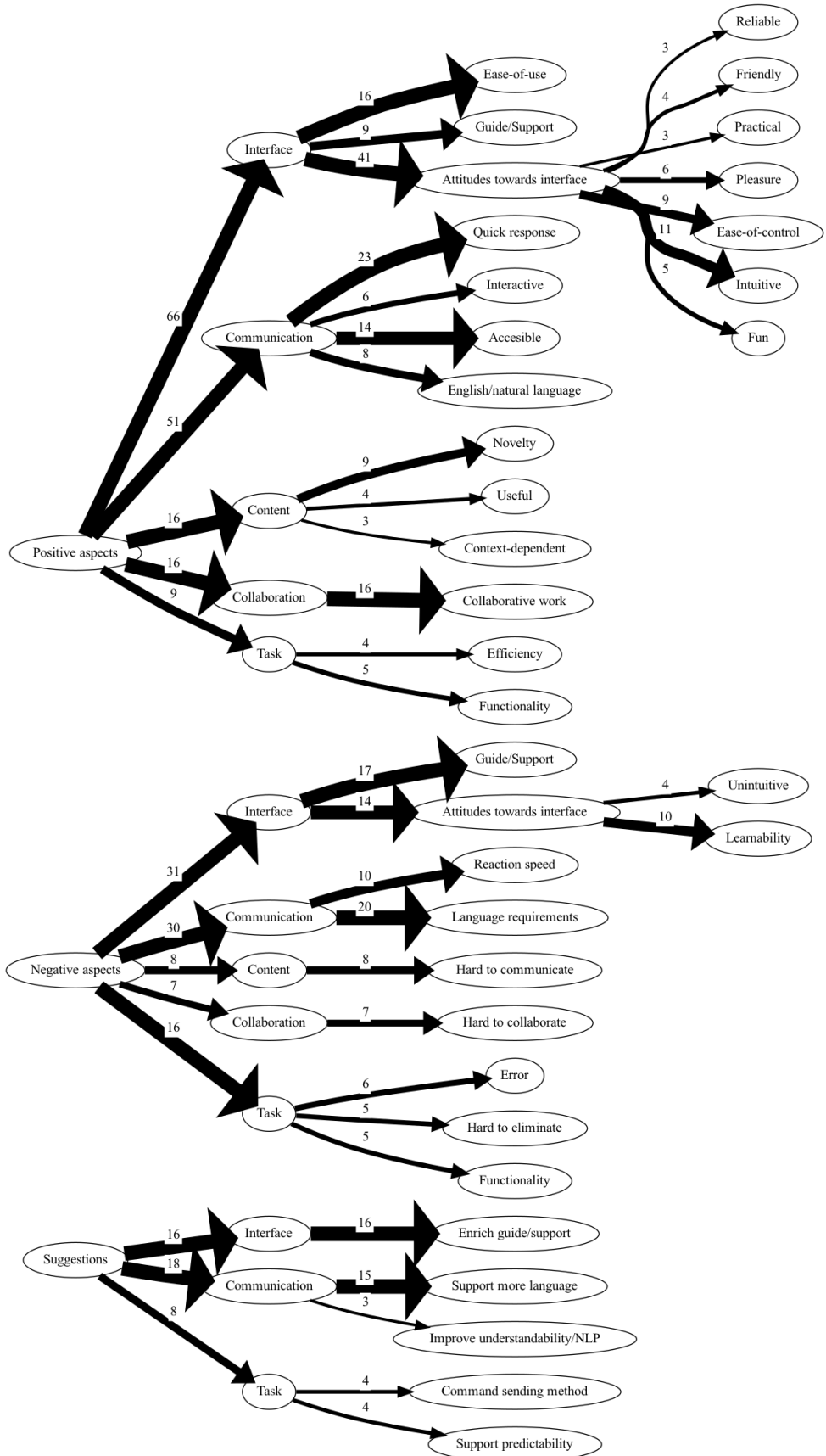

**Figure 2. Thematic analysis graph for SOCIO**

SOCIO received more negative feedback than Creately, where its most frequently reported weakness was that it failed to meet all users' purposes. Of the participants, 30% (20) noted that SOCIO only understands the natural language in English (e.g., "people who do not have a good enough English level could run into difficulties"). Another 26% of participants (17) made the point that SOCIO is not easy to control (e.g., "the use of the tool commands is complicated"). Finally, 21% of participants (14) claimed that SOCIO is not easy to learn (e.g., "it is not that simple to establish relationships, and you have to think a lot about how they should be expressed").

**4.2. Communication and Collaboration**

*Chatbot benefits communication and collaboration efficiency:* Unlike web-based tools, chatbots can be accessed via mobile devices, allowing participants to communicate with each other while modeling. Several participants (14) reported that they could access and edit class diagrams anytime and anywhere using SOCIO. Some (16) also noted that, with SOCIO, they could communicate with teammates while advancing their diagrams. Eight participants mentioned that the SOCIO chatbot was able to understand natural language.

Both collaborative modeling tools strive to be responsive and make updates as quickly as possible. However, the synchronization principles differ. In terms of response speed, participants are more satisfied with SOCIO. Many participants (23) praised SOCIO's fast response on Telegram. All teams observed changes in the class diagram as soon as a command was sent to the chatbot. In contrast, some participants (13) complained that Creately takes time to load and changes are not updated immediately.

*Real-time collaboration is essential for both modeling tools:* Both tools were praised for their ability to support collaborative work in real-time, but SOCIO received less feedback and suggestions than Creately. For instance, 42% of participants using Creately (28) noted that they found it difficult to collaborate in real-time due to difficulties adjusting interdependent actions and non-timely feedback (e.g., "I get things wrong during collaborative work because I don't know what my teammate is doing" and "(Creately) does not let me edit while others are editing"). Creately's bugs and slow response were mentioned more than ten times as negative aspects for supporting real-time work.

**4.3. Content and Task**

There was relatively less negative feedback on either tool for content or tasks. Creately outperforms SOCIO on functionality, as it was praised 12 times more often and did not receive negative feedback. SOCIO's novelty and usefulness appear to surprise users (13 times). However, 16 participants noted that Creately was not easy to use for task performance, and some got confused when starting and sharing projects with teammates. Regarding SOCIO, some participants indicated that removing class diagram elements was troublesome and sometimes buggy, and participants suggested improving command sending and predictive help.

**5. Participant Feedback**

Regarding the three main positive aspects of each compared tool, SOCIO received more positive comments (158) on more aspects than Creately (127). Both tools satisfied participants in collaborative work, ease of use, physical comfort, functionality, pleasure, and quick response. Creately seems to satisfy participants better than SOCIO concerning aspects of collaborative work, ease of use, physical comfort, and functionality. Meanwhile, participants were more satisfied with SOCIO on aspects like quick responses, accessibility, novelty, and provision of help pages.

Both tools received negative comments regarding collaborative work, usage requirements, response speed, and the interface look and feel. Creately received more negative comments (109) than SOCIO (92) about these negative issues.

Several suggestions for improvement were proposed for SOCIO: 24% of the participants indicated that the chatbot should support more languages, and 23% raised the need to improve the help page and online support. In turn, 18% of participants suggested that Creately needs to improve the synchronization of real-time collaborative work. Another 10% thought that Creately should be easier to use. About 9% of the participants requested to improve Creately's GUI. Finally, when asked about their preference, 53% of the participants preferred SOCIO, whereas 47% chose Creately.

## 6. Conclusions

In conclusion, chatbots could achieve similar or higher perceived usability than traditional modeling tools like the one used in this research. As the perceived usability is based on SUS results, the SOCIO chatbot can be said to outperform the Creately web tool quantitatively.

Also, SOCIO appears to outperform Creately qualitatively, as it receives more positive and less negative feedback. Nevertheless, both SOCIO and Creately receive negative reviews, and both tools require some improvements. While chatbots need further improvement regarding providing guidance and help for novices, they appear beneficial for improving user experience.

SOCIO appears helpful in building real-time collaborative models, as it contributes to satisfaction and communication efficiency and improves the humanized user experience. Additionally, natural language communication with chatbots reduces the barriers to modeling for users from different fields. Consequently, SOCIO is potentially a valuable real-time collaboration resource for building a robust real-time modeling tool, especially when collaborators and modeling tools are obliged to interact. We are executing a family of experiments to evaluate satisfaction quantitatively and qualitatively by conducting sentiment analysis and thematic analysis to analyze the experiments and, in turn, increase the reliability of the joint results.

We have identified usability aspects that require improvement in the SOCIO chatbot and, therefore, can be applied to other collaborative chatbots enabling UML modeling. Based on the above analysis, we received many feedback and suggestions. In practice, it is impossible to address all these shortcomings at once. Given that most of the suggestions and criticisms refer to enriching the guidelines and support and supporting more languages, we suggest considering the following features for future collaborative chatbot development:

1. Provide guidance and a help page with the following features: (1) a detailed introduction of the elements accepted by the chatbot, for instance, the attribute types accepted (int, double, float, date, string); (2) command descriptions could include examples of class diagrams, for instance, showing different ways of connecting elements.
2. Provide alternative context-sensitive help by giving multiple responses if the chatbot does not understand users' commands. For instance, when a user sends a wrong command, the chatbot may suggest the correct way to send the command; when a user sends a command with a grammar mistake, the chatbot may automatically correct the mistake.

**Acknowledgment**
This research was funded by the Spanish Ministry of Science, Innovation, and Universities research grants PGC2018-097265-B-I00 and the FINESSE project (PID2021-122270OB-I00). This research was also supported by the Madrid Region R&D program (project FORTE, P2018/TCS-4314) and the SATORI-UAM project (TED2021-129381B-C21).